\documentstyle[11pt]{article}

\begin{document}

\date{}
\title{the Relation Between Technique of Conformal Flat
and Damour-Ruffini-Zhao's Method }
\author{{\ M.X Shao \thanks{{\protect\small E-mail: shaomingxue@hotmail.com}}}
, and {Z. Zhao \thanks{{\protect\small E-mail: blackhole@ihw.com.cn}}} \\
{\small Department of Physics, Beijing Normal University, Beijing
100875, P.R.China.}} \maketitle
\begin{abstract} The relation between the technique of conformal flat
and Damour-Ruffini-Zhao's method is investigated in this paper.
It is pointed out that the two methods give the same results when
the metric has the form $g_{\alpha\beta=0},$ with $\alpha=0,1$ and
$\beta=2,3$. It is indicated  that the two methods are not
equivalent for general case.

PACS:
\newline Keywords:black hole, conformal flat ,temperature
\end{abstract}

\vskip 0.6in

\section{Introduction}
There are some schemes to research Hawking radiation for black
holes\cite{hawking}\cite{D-R}\cite{sannan}\cite{zhao78}\cite{zhao79}
\cite{zhao80}. Zhao  et al. developed an technique to rapidly
determine the location and temperature of event horizon called
conformal flat\cite{zhao95}\cite{zhao96} . Later they applied the
technique to non-stationary black
hole\cite{zhao97}\cite{zhao98}\cite{zhao99}. The technique require
that in tortoise coordinate the metric of two dimensional
subspace $x^0, x^1$ is conformal to two dimensional Minkowski
space. This requirement does simply and rapidly determine the
location and temperature of the horizon simultaneously. This
technique is closely related to Zhao's another
method\cite{zhaobook}\cite{zhao50}\cite{zhao81}\cite{zhao82}\cite{zhao83}\cite{zhao86},
which requires the Klein-Gordon equation has the standard form of
wave equation. Therefore there naturally arise a problem: whether
or not the two schemes give the same results. Our answer is
affirmative for some special case. But for a general case our
answer is negative. In the next section we will investigate this
issue in detail and prove the equivalence between the two schemes
when $g_{\alpha\beta=0},$ with $\alpha=0,1$ and $\beta=2,3$.  In
section three we illustrate that the two methods is not
equivalent for general black hole.

\section{the Special Case  the Two Scheme Is Equivalent}

Consider the case that the metric is
\begin{equation}\label{c0}
g_{\mu\nu}=\left[
\begin{array}{cccc}
  g_{00} & g_{01} & 0 & 0 \\
  g_{10} & g_{11} & 0 & 0 \\
  0      & 0      & g_{22} & g_{23} \\
  0      & 0      & g_{32} & g_{33}
\end{array}\right]
\end{equation}
ie
\begin{equation}\label{c1}
g_{\alpha\beta}=0,~~ with~~ \alpha=0,1 ~~and~~ \beta=2,3.
\end{equation}

For the horizon $\xi=\xi(x^0)$, the tortoise transformation is
\begin{equation}\label{c2}
x^1_*=x^1+\frac{n}{2\kappa}ln(x^1-\xi)
\end{equation}
with other components of coordinates invariant.  Therefore in the
new coordinates $(x^0,x^1_*,x^2,x^3)$,

\begin{equation}\label{c3}
\tilde{g_{\alpha\beta}}=\frac{\partial x^{\lambda}}{\partial
\tilde{\alpha}}  \frac{\partial x^{\rho}}{\partial \tilde{\beta}}
g_{\lambda\rho} =g_{\alpha\beta}=0,
\end{equation}
where $\alpha=0,1$ and $\beta=2,3$ and (\ref{c1})(\ref{c2}) has
been used to obtain the Eq. (\ref{c3}). Therefore only  $0,1$
components of metric change.
\begin{equation}\label{c5}
\tilde{g_{\mu\nu}}=\left[
\begin{array}{cccc}
  \tilde{g_{00}} & \tilde{g_{01}} & 0 & 0 \\
  \tilde{g_{10}} & \tilde{g_{11}} & 0 & 0 \\
  0      & 0      & \tilde{g_{22}} & \tilde{g_{23}} \\
  0      & 0      & \tilde{g_{32}} & \tilde{g_{33}}
  \end{array} \right].
\end{equation}

In the two-dimensional subspace $(x^0, x^1_*)$,
\begin{equation}\label{c6}
ds^2=\tilde{g_{00}}dx^0dx^0+\tilde{g_{11}}dx^1_*dx^1_*+ other~~ terms\\
=\tilde{g_{00}}[dx^0dx^0+\frac{\tilde{g_{11}}}{\tilde{g_{00}}}dx^1_*dx^1_*]+other
~~terms.
\end{equation}

The technique of conformal flat require
\begin{equation}\label{c7}
\frac{\tilde{g_{11}}}{\tilde{g_{00}}}=-1,
\end{equation}
While he Damour-Ruffini method gives the condition as
\begin{equation}\label{c8}
  \frac{\tilde{g^{11}}}{-\tilde{g^{00}}}=1.
\end{equation}

From(\ref{c5}), one can easily obtain
\begin{eqnarray}\label{c9}
\tilde{g^{00}}=\frac{\tilde{g_{11}}}{\tilde{g_s}}\\
\tilde{g^{11}}=\frac{\tilde{g_{00}}}{\tilde{g_s}},
\end{eqnarray}
where $\tilde{g_s}$ is the determinant of submatrix of (\ref{c5})
\begin{equation}
\tilde{g_{s}}=\left|
\begin{array}{cc}
  \tilde{g_{00}} & \tilde{g_{01}}  \\
  \tilde{g_{10}} & \tilde{g_{11}}  \\
  \end{array} \right|.
\end{equation}
At once Eq.(\ref{c5}) ensure the equivalence between Eq.(\ref{c7})
and Eq.(\ref{c8}).

Next we investigate the equivalence  in Eddington coordinates
\begin{equation}\label{c11}
  ds^2=g_{00}dx^0dx^0+2g_{01}dx^0dx^1+g_{11}dx^1dx^1+ (other~~
  terms).
\end{equation}

After take tortoise coordinate transformation, ones obtain  in
the new coordinate $ds^2$
\begin{equation}\label{c12}
  ds^2=\tilde{g_{01}}\{\frac{\tilde{g_{00}}}{\tilde{g_{01}}}dx^0dx^0+
  2dx^0dx^1_*\} + other~~ terms.
\end{equation}

The generalized conformally flatization requires
\begin{equation}\label{c13}
\frac{\tilde{g_{00}}}{\tilde{g_{01}}}=-1.
\end{equation}

In the other hand,
\begin{equation}\label{c14}
  \tilde{g^{00}}=\frac{\tilde{g_{11}}}{\tilde{g_s}},~~~
\tilde{g^{01}}=-\frac{\tilde{g_{01}}}{\tilde{g_s}}.
\end{equation}

The Klein-Gordon equation in $x^1_*, x^0$
\begin{equation}\label{c15}
  \frac{\tilde{g}^{11}}{\tilde{g}^{01}}  \frac{\partial^2}{(\partial
  x^1_*)^2}\Phi+2\frac{\partial^2}{\partial x^0x^1_*}\Phi+
  (others)=0.
\end{equation}

 Damour-Ruffini-Zhao's method require the coefficient of
 $\frac{\partial^2}{(\partial x^1_*)^2}$ being 1
\begin{equation}\label{c16}
 \frac{\tilde{g^{11}}}{\tilde{g^{01}}}=1.
\end{equation}
From Eq.(\ref{c14}) and Eq.(\ref{c16}), Eq.(\ref{c13}) is obtained
at once. Therefore in Eddington coordinates, the technique of
conformal flat is equivalent to Damour-Ruffini-Zhao's method.

\section{Technique of Conformal Flat in General Metric}

We will prove that there does not exist the equivalence for
general case.

In the previous section, we focus on the special case that the
event horizon $\xi=\xi(x^0)$. In this subsection we analyze the
general case that $\xi=\xi(x^0,x^2,x^3)$ in Eddington coordinates.

Suppose $x^1=\xi (x^0,x^2,x^3),$.  Tortoise coordinates
transformation,
\begin{equation}
x_{*}^1=x^1+\frac1{2\kappa }ln(x^1-\xi ) \label{e4}
\end{equation}
with other components invariant.
\begin{equation}\label{c17}
dx^1_*=(1+\frac{1}{\epsilon})dx^1-
\frac{\xi^{\prime}_{\nu}}{\epsilon}dx^{\nu},
\end{equation}
in which $\epsilon=2\kappa(x^1-\xi)$ and
$\xi^{\prime}_\nu=\frac{\partial \xi}{\partial x^\nu}$.

 The metric is then obtained in terms of tortoise coordinates
\begin{equation}\label{c18}
ds^2=(\frac{\epsilon g_{11}\xi^{\prime}_0}{(1+\epsilon)^2}+
\frac{\epsilon g_{10}}{1+\epsilon})
[\frac{\frac{g_{11}}{(1+\epsilon)^2}\xi^{\prime}_0\xi^{\prime}_0
+\frac{2g_{10}}{1+\epsilon} \xi^{\prime}_0+g_{00}}{
\frac{\epsilon g_{11}}{(1+\epsilon)^2}\xi^{\prime}_0+
g_{10}\frac{\epsilon}{1+\epsilon}}dx^0dx^0+2dx^0dx^1_*]+(others).
\end{equation}

The technique of conformally flat require the coefficient of
$dx^0dx^0$ in $[~]$ in Eq.(\ref{c18}) being $-1$
\begin{equation}\label{c20}
\frac{\frac{g_{11}\xi^{\prime}_0\xi^{\prime}_0+2g_{10}\xi^{\prime}_0+g_{00}}
{\epsilon}+2g_{00}+2g_{10}\xi^{\prime}_0+\epsilon g_{00}}
{g_{10}+g_{11}\xi^{\prime}_0+\epsilon g_{10}}=-1.
\end{equation}

When $x^1 \mapsto \xi, \epsilon \mapsto 0$, the well-definition of
the numerator of lhs. of Eq.(\ref{c20}) deduces
\begin{equation}\label{c21}
g_{11}\xi^{\prime}_0\xi^{\prime}_0+2g_{10}\xi^{\prime}_0+g_{00}=0.
\end{equation}

Now we show that the  event horizon determined by (\ref{c21}) is
not equivalent to that given by Damour-Ruffini-Zhao's
method\cite{gr-qc0010078}, 
which determines the location of horizon as
\begin{equation}\label{c21a}
g^{11}-2g^{1\nu }\xi _\nu ^{\prime }+g^{\mu \nu }\xi _\mu
^{\prime }\xi _\nu ^{\prime }=0.
\end{equation}

Considering the special case that $\xi^{\prime}_0=0$,
Eq.(\ref{c21}) simplifies to
\begin{equation}\label{c21b}
g_{00}|_{x^1\mapsto \xi}=0
\end{equation}

But when $\xi^{\prime}_0=0$, Eq.(\ref{c21a})
simplifies
\begin{equation}\label{c22}
g^{11}-2g^{1\nu }\xi _\nu ^{\prime }|_{\nu=2,3}+g^{\mu \nu }\xi
_\mu ^{\prime }\xi _\nu ^{\prime }|_{\mu\nu=2,3}=0.
\end{equation}

Clearly Eq.(\ref{c21}) is not equivalent to Eq.(\ref{c22})
generally. Therefore the event horizons determined  by the two
equations are not the same in general case. Apparently the
$\kappa$'s are different either.

The reason is that  Damour-Ruffini-Zhao's method uses $g^{\mu\nu}$
in generalized tortoise coordinates to determine $\xi$ and
$\kappa$, while technique of conformal flat uses $g_{\mu\nu})$ in
generalized tortoise coordinates  to determine them. Generally
speaking, they do not produce the same results.

At a first glance it seems that kerr  metric with term $dx_0dx_3$
is an exception that our investigation in section two does not
include. But when the metric of  Kerr black hole is written  in
the dragging system\cite{zhao1}\cite{zhao2}, then it is included
in our investigation.

\end{document}